# A zero-carbon, reliable and affordable energy future in Australia

Bin Lu*, Andrew Blakers, Matthew Stocks, Cheng Cheng and Anna Nadolny
Australian National University
*Correspondence: bin.lu@anu.edu.au

## Abstract

Australia has one of the highest per capita consumption of energy and emissions of greenhouse gases in the world. It is also the global leader in rapid per capita annual deployment of new solar and wind energy, which is causing the country's emissions to decline. Australia is located at low-moderate latitudes along with three quarters of the global population. These factors make the Australian experience globally significant. In this study, we model a fully decarbonised electricity system together with complete electrification of heating, transport and industry in Australia leading to an 80% reduction in greenhouse gas emissions. An energy supply-demand balance is simulated based on long-term (10 years), high-resolution (half-hourly) meteorological and energy demand data. A significant feature of this model is that short-term off-river energy storage and distributed energy storage are utilised to support the large-scale integration of variable solar and wind energy. The results show that high levels of energy reliability and affordability can be effectively achieved through a synergy of flexible energy sources; interconnection of electricity grids over large areas; response from demand-side participation; and mass energy storage. This strategy represents a rapid and generic pathway towards zero-carbon energy futures within the Sunbelt.

Keywords: solar photovoltaics; wind energy; energy security; energy storage; Super Grid; Smart Grid

## 1. Introduction

Solar photovoltaics and wind turbines comprised about 64% of global annual net new capacity additions in 2019 [1], and nearly 100% in Australia. Rapid deployment of solar and wind energy presents the most promising prospect for tackling climate change through the adoption of renewable energy in the electricity sector, along with electrification of heating, transportation and industry to displace fossil fuels. However, solar and wind energy are weather-based and hence are variable and uncertain in nature. Consequently, there are a range of technical challenges associated with the large-scale integration of variable renewable energy such as higher ramp rates, lower minimum generation levels, more frequent cycling and capacity inadequacy.

Energy storage is key to a reliable and affordable renewable energy future. Jacobson et al. [2, 3] modelled thermal energy storage to support 100% wind, water and sunlight in the United States and the world's energy systems. Phase-change materials were included to store high-temperature heat from concentrated solar power, which was then used to drive steam turbines to generate electricity when needed. Hot water, chilled water, ice and underground rocks were used to store low-temperature heat from solar thermal collectors and electricity to meet heating and cooling demand for those times when energy supply and demand were not balanced. Demand response was also modelled where 15% of residential and commercial, 85% of transport and 70% of industrial loads were assumed to be flexible – providing up to 8 hours of load shifting. Connolly et al. [4] and Lund et al. [5] investigated large-scale integration of solar and wind energy in Europe using a Smart Energy System solution. The electricity, heating, cooling and transport sectors were coupled through power-to-gas, where solar and wind energy were used to produce methane, methanol and dimethyl ether mainly for transport fuels, but also for electricity and heat generation. In this way, variable renewable energy can be stored in the form of electrofuels in gas and oil storage facilities, which are largely available today at low cost. Additionally, instead of being burned directly, biomass was utilised as a carbon source to produce bio-electrofuels using gasification and hydrogenation processes. Ram et al. [6] and Bogdanov et al. [7] modelled the energy transition required to decarbonise global power, heat, transport and desalination. Lithium-ion batteries were used for short-term energy storage i.e. energy day-night shifting. Power-to-gas and compressed air energy storage were utilised for medium-term to long-term energy storage to cope with seasonal variations of renewable energy resources. About 5% of electricity demand and more than 10% of heat demand were powered by synthetic natural gas through power-to-gas. Further, about 15% of heat demand was met by thermal energy storage, including industrial heat (medium- to high-temperature) and space and water heating (low-temperature).

In this study, by contrast, we address the variability and uncertainty of renewable energy in a different way, using short-term off-river energy storage (STORES) and distributed energy storage (DES). STORES refers to closed-loop pumped hydro systems with large hydraulic head, which can be located away from rivers and hence offers vast opportunities to access cost-effective mass energy storage. A first-of-its-kind high-resolution global atlas of off-river pumped hydro included in Blakers et al. [8] demonstrated 616,000 cost-effective sites for pumped hydro development around the world with a total storage potential of 23 million GWh. DES, such as electric car batteries, can contribute significant storage capacity as well as large demand flexibility to future energy systems. Enabled by smart grid technology, these kW, kWh-scale storage systems can be aggregated and utilised for GW, GWh-scale energy storage. In light of their high round-trip efficiencies (STORES 80%, DES 90%) and the large resource potentials, these two solutions are ideal for short-term, diurnal energy storage.

A set of 100% renewable electricity futures in Australia are modelled in this work. Australia has one of the highest greenhouse gas emissions per capita and is the largest exporter of coal (#1) and liquified natural gas (#2) in the world. However, Australia is a global leader in rapid per-capita deployment of renewable energy as shown in Fig. 1: over the years 2018-2020, the combined solar photovoltaics and wind deployment will be above 15 GW, which is greater than 200 watts per capita per year – 4 times the per capita rate for the European Union, the United States, China and Japan and 10 times the global average [8]. If this rate were to continue, Australia would be on track for 50% renewable electricity in 2025 and 100% in the early 2030s [9]. The modelling of a zero-carbon renewable electricity future makes a timely contribution to ongoing discussions on energy security and affordability. Importantly, about three quarters of the world’s population lives in the “Sunbelt” (lower than 35 degrees of latitude), where the solar irradiance is high, the seasonal variation in solar resource is low, and there is no significant heating load in winter. Therefore, long-term, seasonal energy storage requirements are low compared with Europe, North America and Northeast Asia. Most of the Sunbelt countries can readily follow the Australian path, transitioning to a high renewable energy future and bypassing a fossil fuel era [8].

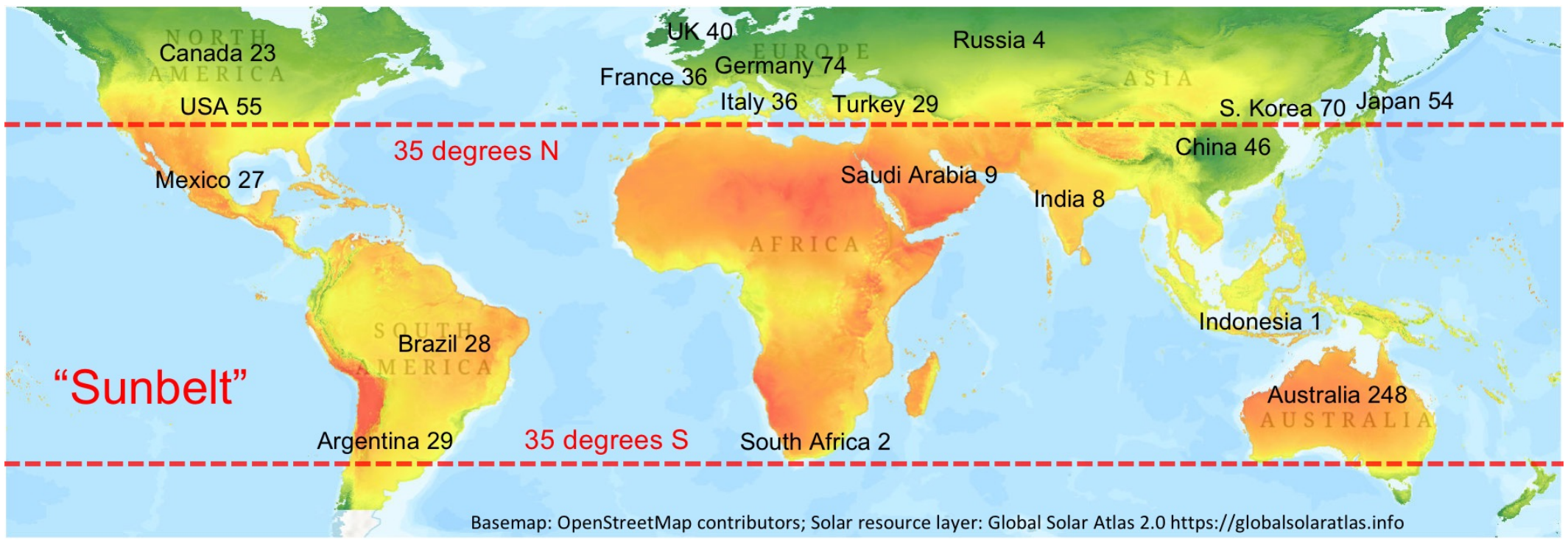

Fig. 1: Renewable energy deployment rates (watts per person per year) in 2019. Data source: International Renewable Energy Agency [1]. Green to red colours denote the daily average Global Horizontal Irradiation ranging from 1.3 to 7.5 kilowatt-hours per square metre. Data source: Solargis [10].

## 2. Methodology

Energy generation, storage and transmission were simulated in a set of three 100% renewable electricity scenarios for Australia on a 30-minute basis over the years 2020-2029. Within all three scenarios, the electrical energy demand included the current electricity demand in the electricity sector together with electrified land transport, heating, manufacturing and mining. Powered by renewable energy, this represents a reduction of 80% in total Australian greenhouse gas emissions, which currently amount to 532 megatonnes $CO_2$-e [11] or 21 tonnes per person. As part of this 80% reduction, fugitive emissions from Australia's exports of coal and gas were also assumed to be eliminated.

Scenario 1: "7 Grids". This is the baseline scenario in which we assumed that regional electricity markets were operated separately in 7 Australian states and territories: New South Wales (NSW), Northern Territory (NT), Queensland (QLD), South Australia (SA), Tasmania (TAS), Victoria (VIC) and Western Australia (WA). In other words, each state/territory (sub-scenarios 1.1-1.7) transitioned to a zero-carbon electricity future in its own way: for example, hydropower played a significant role in the island state of TAS, while solar photovoltaics (PV) and wind constituted the majority of the energy mix in the mainland states/territories of NSW, NT, QLD, SA, VIC and WA. Energy storage, in the form of short-term off-river energy storage, supported high penetration of variable solar and wind energy through large-scale energy time-shifting, and a variety of ancillary services, such as frequency control and black start capability. This scenario reflects the status quo of the existing Australian electricity systems, which are weakly interconnected and isolated from electricity networks in neighbouring countries such as Indonesia and New Zealand.

Scenario 2: "Super Grid". In this scenario, energy systems in NSW, QLD, SA, TAS and VIC were fully integrated as a National Electricity Market (NEM), along with 3 potential extensions (sub-scenarios 2.1-2.8) to Far North Queensland (FNQ, 1,500 km), NT/Alice Springs (1,200 km) and WA/Perth (2,400 km). As shown in Fig. 2, a high-voltage direct-current (HVDC) backbone was envisaged on top of the existing transmission network, connecting widely dispersed renewable energy zones across the Australian continent. While energy storage is still critical in the Super Grid scenario, the storage requirements were expected to be significantly reduced due to the smoothing effect of less correlated and sometimes anti-correlated renewable energy resources over large areas.

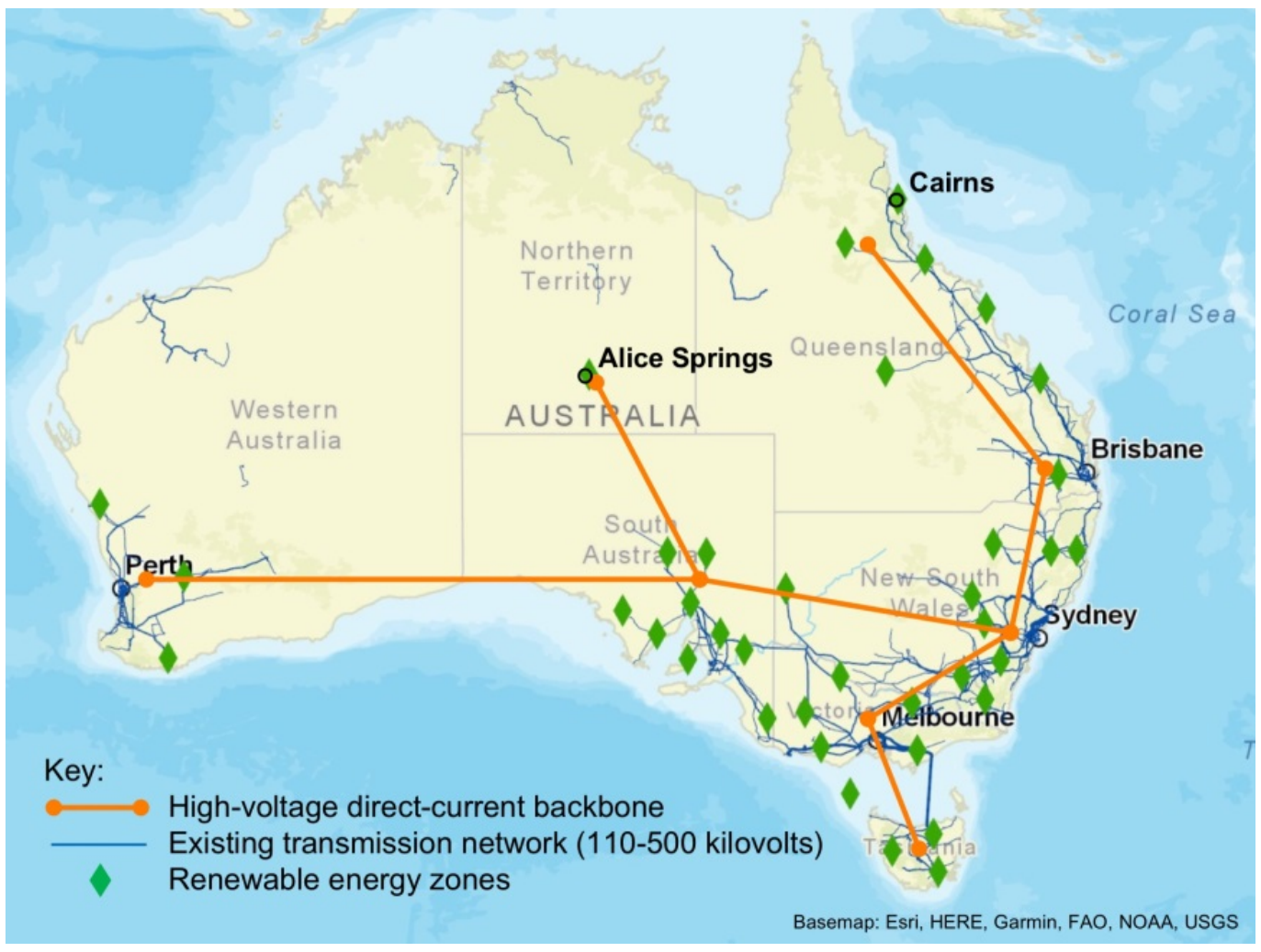

Fig. 2: A hypothetical high-voltage direct-current backbone (orange) lies on top of existing transmission network (blue). Renewable energy zones are connected to the adjacent transmission nodes by high-voltage alternating-current transmission lines.

Scenario 3: “Smart Grid”. This scenario (comprising sub-scenarios 3.1-3.8) was built based on the Super Grid scenario, with an additional assumption that distributed energy storage such as electric car batteries contributed large demand flexibility to the electricity system, enabled by smart grid technology. In the modelling, 80% of passenger cars were assumed to be compatible with flexible charging in response to energy deficits or surpluses in the electricity system, subject to a minimum state-of-charge level of 25%. This scenario represents a promising future for active demand-side participation in the energy market. Note that no electricity was drawn from the vehicle batteries into the grid (no “vehicle to grid”) in this scenario; rather, the charging times are managed.

The modelling framework is illustrated in Fig. 3. The modelling inputs and assumptions are discussed in Sections 2.1-2.5.

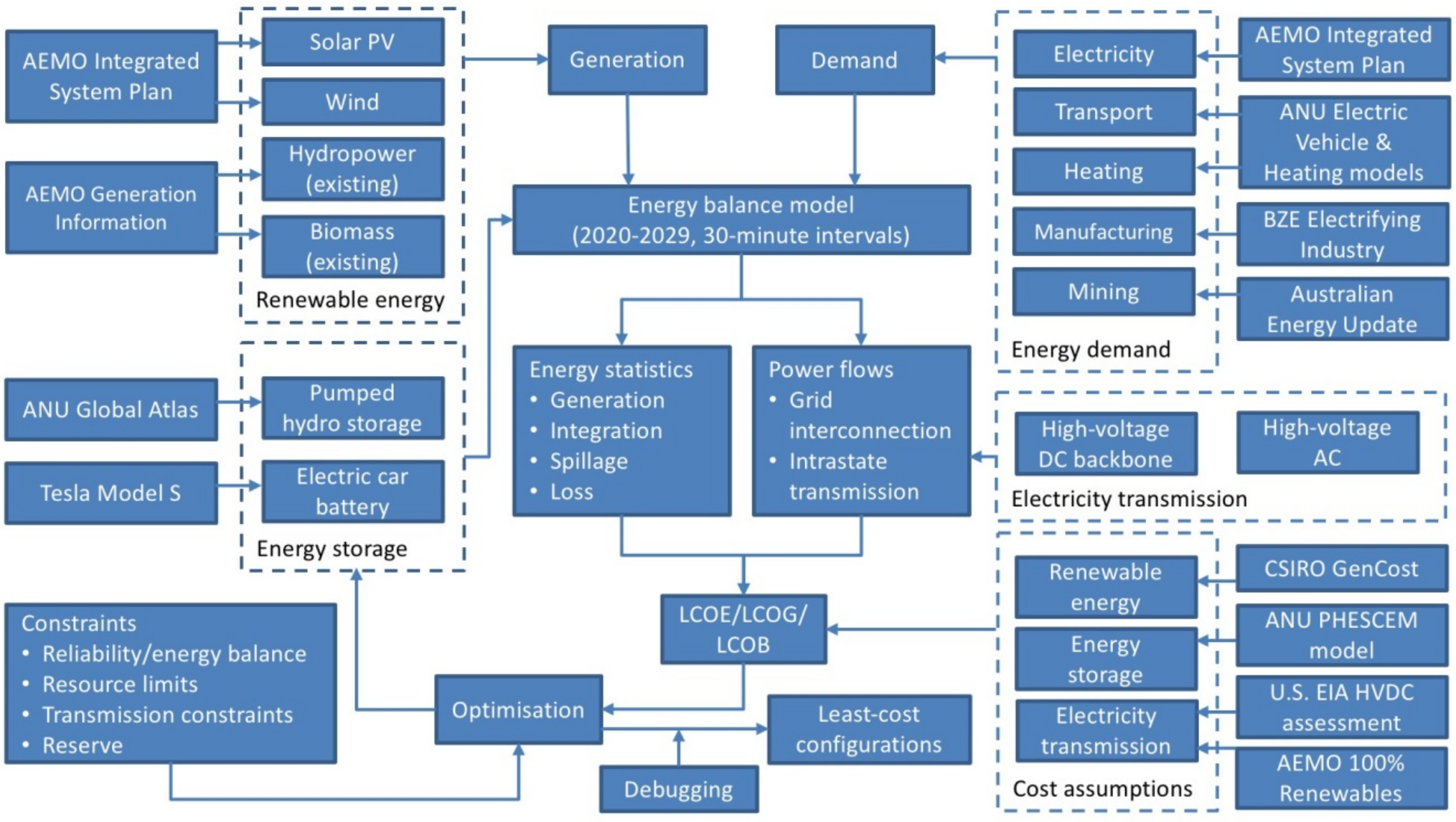

Fig. 3 Modelling framework: the nucleus of the model is a high-resolution analysis of energy supply and demand balance based on long-term, chronological meteorological and energy demand data. Differential Evolution is utilised to optimise configurations of electricity generation, storage and transmission technologies. Acronyms and abbreviations: Australia Energy Market Operator (AEMO), Australian National University (ANU), Beyond Zero Emissions (BZE), Commonwealth Scientific and Industrial Research Organisation (CSIRO), U.S. Energy Information Administration (U.S. EIA).

### 2.1 Renewable energy

In this study, solar and wind energy were the major energy sources (> 90%) in the renewable electricity systems, with support provided from existing (but no new) hydropower and biomass. High-resolution (30-minute) solar and wind energy traces for 2020-2029 were obtained from the Integrated System Plan 2018 developed by AEMO [12]. The Integrated System Plan detailed a transition pathway for the Australian National Electricity Market in the coming decades and included a set of projected solar and wind energy time series for each renewable energy zone in Fig. 2, which are publicly available. For WA and NT, which are not connected to the National Electricity Market and hence were not covered by the report, the 2008-2017 meteorological data from the Australian Bureau of Meteorology [13, 14] were used and "downscaled" to 30-minute intervals by interpolation where required. The methodology of solar and wind energy conversions was described in Section 4.1, 4.2 of Lu et al. [15]. On average, the capacity factors across the renewable energy zones are 30% AC for solar PV with single-axis trackers and 41% for onshore wind, respectively.

For other renewables such as existing hydro and bio, it was assumed that they would stay unchanged from the current level (no further expansion) and were fully dispatchable throughout the simulated period, subject to current energy and power constraints. Historically, the annual electricity generation from existing hydro and bio ranged from 15-22 TWh since 2000 [16, 17]. Thus, the contribution of hydro and bio was constrained to a maximum of 20 TWh per year in the modelling. Future opportunities for significant expansion of river-based hydropower [18] are small compared with the massive scale of solar PV and wind required to reach 100% renewable electricity. Intensive use of bioenergy, whether by the burning of biomass or utilization of biofuels, would contribute to significant air pollution and increased ozone-related health risks. Additionally, large-scale biomass competes with food, forests and ecosystems for land, water, fertilisers and pesticides [19].

Nuclear energy is not included in this study. Nuclear energy is associated with public perceptions of weapons proliferation, accidents and waste disposal. Furthermore, the nuclear industry has a low growth rate in terms of global net new generation capacity: an average of 2.2 GW per year over the past decade [20].

### 2.2 Energy storage

Energy storage makes energy time-shifting possible, and also provides a variety of ancillary services, such as frequency regulation, which can facilitate large-scale integration of solar and wind energy in an electricity system. Large-scale energy storage technologies include pumped hydro, high-temperature thermal (power-to-heat), grid-scale battery, compressed air, electrolytic hydrogen and renewable electrofuels (power-to-gas).

Compared with alternative energy storage technologies, pumped hydro has low costs, by far the highest technology maturity, and a high round-trip efficiency of typically 80%. Indeed, because of these advantages, pumped hydro constitutes 97% (rated power) or 99% (storage capacity) of the global energy storage market [21]. However, in many parts of the world, hydro energy resources are extremely limited and hence opportunities for further developments of large-scale river-based hydroelectric projects in these regions are restricted. In addition, developments of significant hydroelectric schemes are usually associated with a wide variety of environmental concerns such as biodiversity, nutrient flows and landscape destruction.

By contrast, short-term off-river energy storage (STORES), which refers to closed-loop pumped hydro systems with large hydraulic head (typically > 300 m), can be located away from rivers and hence offers vast opportunities to access cost-effective mass energy storage. A “first-of-its-kind” comprehensive global atlas of pumped hydro (off-river) has been developed at the Australian National University [8], which discovered 616,000 cost-effective sites for pumped hydro energy storage development around the world – 3,000 of them are located in Australia. A large difference in altitude between upper and lower reservoirs enables significant amounts of energy to be stored in pairs of medium-sized closed-loop reservoirs. Closed-loop systems mean there is no or low interaction with the ecosystem of main stem rivers, which reduces impacts on the environment and natural landscape. The consumption of water is modest (initial fill and evaporation minus rainfall). In this study, STORES is included in the modelling and utilised for large-scale renewable energy time-shifting and energy demand balancing.

As well as large-scale energy storage, small-scale distributed energy storage (DES) systems, such as electric car batteries, were also included in the modelling. Enabled by smart grid technology, these kW, kWh-scale DES systems could be aggregated and utilised for effective GW, GWh-scale demand response to mitigate energy and power deficiency due to occasional low availability of renewable energy. In this study, the charging of 80% of the passenger cars was assumed to be fully flexible and regulated according to a real-time energy supply and demand balance, while subject to a minimum state-of-charge constraint of 25%.

### 2.3 Electricity transmission

In addition to energy storage (time-shifting), wide geographic dispersion of solar and wind resources can also effectively mitigate intermittency in energy production and consumption (i.e. energy geo-shifting). Renewable energy resources and electricity demand are generally less correlated or even anti-correlated over a large geographic area e.g. one million square kilometres of land in Australia’s east coast National Electricity Market.

Modern high-voltage direct-current (HVDC) technology, with either line-commutated converters or voltage-source converters, enables cost-effective delivery of GW-scale electric power over thousands of kilometres with relatively low transmission loss (3% per 1,000 km). In this study, as shown in Fig. 2, a hypothetical HVDC backbone was envisaged which spreads across the widely dispersed renewable energy zones and lies on top of the existing transmission network. It was utilised for large-scale export of renewable energy from FNQ/Cairns, NT/Alice Springs and WA/Perth to the National Electricity Market and for stronger interconnection between the electricity grids in NSW, QLD, SA, TAS and VIC, which are currently weakly interconnected.

However, we note that while HVDC technology has the advantage of long-distance bulk power transmission at moderate cost, there is a risk that even a single-pole transmission failure could lead to loss of GW-scale electric power, which may cause severe capacity inadequacy in the system. Therefore, for both DC transmission lines and DC/AC converter stations, an “N-1” redundancy was applied in the cost assumptions, which incorporated 25% reserve capacity in a two-circuit bipolar HVDC transmission route. Additionally, the renewable energy zones were assumed to be connected to adjacent HVDC nodes by high-voltage alternating-current (HVAC) transmission. The capital costs of both the HVDC and HVAC were factored into the cost calculation, which is a critical component as demonstrated in the results.

## 2.4 Energy demand

The modelled operational demand (excluding behind-the-meter rooftop solar) was 397 TWh p.a. on average. This included energy demand in the current electricity sector and the fully electrified energy consumption for residential & commercial, manufacturing, mining and land transport, as shown in Fig. 4. This transition would allow for an 80% reduction in total Australian greenhouse gas emissions. Thus, the historical electricity demand in the NEM (203 TWh) was doubled.

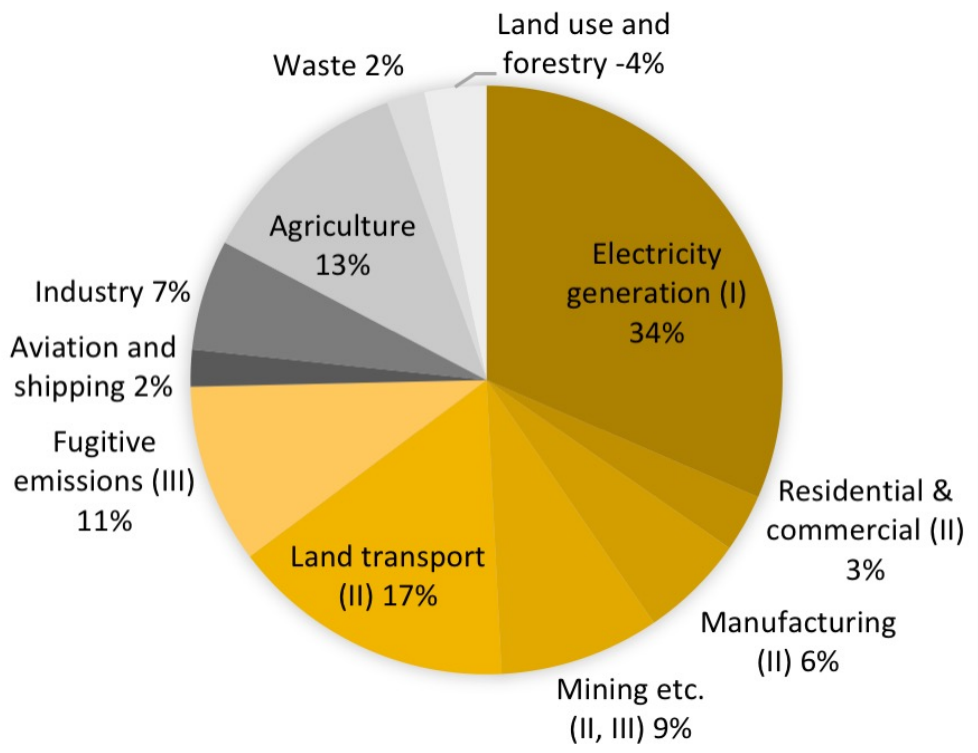

| Category | Pathway towards deep decarbonisation of energy sector | |
|---|---|---|
| I | Renewable energy | ▪ Power generation: solar photovoltaics, wind, existing hydro and bio<br>▪ Energy storage: pumped hydro, batteries and thermal energy storage<br>▪ Electricity transmission: high-voltage direct-current and alternating-current |
| II | Direct electrification | ▪ Heating: electric heat pumps and appliances, electric arc/resistance furnaces<br>▪ Transportation: electric vehicles<br>▪ Industry: electric mining and construction equipment |
| III | Vanishing as fossil fuels phased out | |

Fig. 4: Australia’s greenhouse gas emissions by sector in 2018-19 and the pathway towards deep decarbonisation of energy sector. Data source: Australian Department of the Environment and Energy [11]; emissions breakdown based on the National Greenhouse Gas Inventory [22].

Similar to the solar and wind energy traces, energy demand (30-minute time series) in the electricity sector in the NEM was obtained from AEMO [12] for 2020-2029, with the assumptions that economic growth and the future uptake of distributed energy technology are moderate (neutral). For WA/Perth, the historical electricity data from 2008-2017 were used [23]. The average annual energy demand in the original electricity sector is 201 TWh compared with about 203 TWh in 2017-18, which reflects a flat operational consumption over the simulated period.

Land transport and heating (including space heating, water heating and cooking) data were obtained from the Australian National University electric vehicle and electric heating models. The total energy use in 2017-18 for transport and heating [24] was utilised as a benchmark. Energy consumption patterns for passenger cars, light commercial vehicles, rigid trucks, articulated trucks, non-freight trucks, buses and motorcycles were derived from publicly available sources as noted in Table SI-1 in the Supplementary Information. The heating load profiles were calculated based on the temperature, occupancy and historical usage profiles in residential and commercial buildings.

Overall, electrification of land transport and heating resulted in a 58% increase (transport 48%, heating 10%) to the original electricity demand. Tables SI-1, SI-2 in the Supplementary Information include a summary of the assumptions for the modelling of electrified land transport and heating.

Industrial loads, such as manufacturing, mining and construction, were derived from the Australian Energy Update and Beyond Zero Emissions reports [24, 25] with the overall fuel efficiency boosted by a factor of 2 using electric arc/resistance furnaces for heating, and electric mining and construction equipment for motive energy. A flat 24/7 electricity consumption pattern was assumed, which translated to a continuous electricity load of 9 GW. A significant off-grid industrial centre, Mt Isa located in Far North Queensland, was connected to the renewable energy zones as well as the HVDC node. Electrification of manufacturing, mining and construction contributed another 40% increase in the original electricity demand, bringing the total increase to the original electricity load to 98%.

### 2.5 Cost assumptions

Cost assumptions for electricity generation, storage and transmission technologies were included in Table 1. The costs quoted are in Australian dollars, which has a value of about US$0.7. A nominal discount rate of 6.5% was assumed in the cost calculation to reflect the integrated rates for the returns on investment (30% of the capital with a 10% internal rate of return) and the interest rates from banks (70% of the capital with a loan interest rate of 5%). This translated to a real discount rate of 5% by factoring in an inflation rate of 1.5%.

Table 1. Cost assumptions for electricity generation, storage and transmission technologies. Data source: Graham et al. [26], Blakers et al. [27], EIA [28], Tamblyn [29], AEMO [30]

| Technology | Capital cost (A$/kW) | Fixed O&M cost (A$/kW-year) | Variable O&M cost (A$/MWh) | Lifetime (years) |
|---|---|---|---|---|
| Photovoltaics (1-axis tracking) | 1,200 (DC)<br>1,600 (AC) | 15 (DC)<br>18 (AC) | 0 | 25 |
| Wind (onshore) | 1,800 | 36 | 3 | 25 |
| Pumped hydro (off-river) | 800 / 70 [a] | 10 | 0 | 50 |
| High-voltage direct-current (overhead) | 320 / 160,000 [b] | - [c] | 0 | 50 |
| High-voltage direct-current (submarine) | 4,000 [d] | - [c] | 0 | 50 |
| High-voltage alternating-current | 1,500 | - [c] | 0 | 50 |

Note.
[a] $800/kW for power components including turbines, generators, pipes and transformers; plus $70/kWh for storage components such as dams, reservoirs and water
[b] $320/MW-km for transmission lines; plus $160,000/MW for converter stations
[c] Transmission operating and maintenance costs are not included in the LCOE calculation
[d] Including submarine cables and converter stations

In Australia, the levelised costs of solar PV and wind energy have already been in the range of $50-65/MWh and continue to fall rapidly [9]. The figures for solar PV and wind in Table 1 are equivalent to a levelised cost of electricity of $50/MWh for both solar PV and wind in the 2020s, assuming an average capacity factor of 30% AC for solar PV and 41% for wind across the renewable energy zones. Similarly, existing hydro and bio were assumed to be available at a purchase price of $50/MWh, rather than merely factoring in their operating and maintenance (O&M) costs. As noted in Section 2.3, an “N-1” redundancy (25% reserve capacity) was included in the HVDC cost.

## 3. Results

The modelling results are shown in Fig. 5. Details of energy generation, storage and transmission information are included in Table 2.

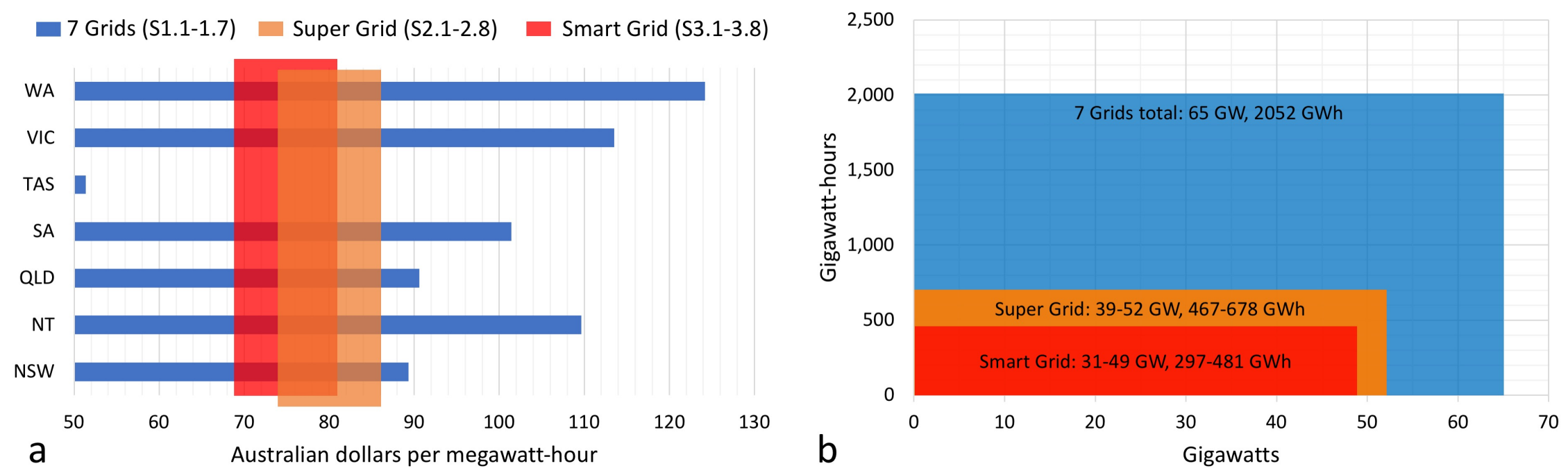

Fig. 5: Levelised costs of electricity (a) and storage requirements (b) in the 7 Grids (blue), Super Grid (orange) and Smart Grid (red) scenarios. The volume-weighted average of LCOE in the 7 Grids scenario is $99/MWh.

Table 2. Rated power (GW), storage capacity (GWh), the annual average of energy production and consumption (TWh), and cost ($/MWh) information for each scenario.

| | Scenario | Energy demand (TWh) | HVDC loss (TWh) | Solar PV | | Wind | | Hydropower | | Biomass | | Pumped hydro | | High-voltage direct-current (GW) | | | | | | | LCOE ($/MWh) | LCOG ($/MWh) | LCOB ($/MWh) | | |
|---|---|---|---|---|---|---|---|---|---|---|---|---|---|---|---|---|---|---|---|---|---|---|---|---|---|
| | | | | GW | TWh | GW | TWh | GW | TWh | GW | TWh | GW | GWh | FNQ-QLD | NSW-QLD | NSW-SA | NSW-VIC | NT-SA | SA-WA | TAS-VIC | | | Storage | Transmission | Spillage & loss |
| 7 Grids | S1.1 NSW | 116 | 0 | 42 | 88 | 15 | 55 | 2 | 2 | 0.1 | 0.1 | 19 | 531 | - | - | - | - | - | - | - | 89 | 49 | 27 | 1 | 12 |
| | S1.2 NT (Alice Springs) | 0.3 | 0 | 0.1 | 0.3 | 0.1 | 0.2 | 0.0 | 0.0 | 0.0 | 0.0 | 0.1 | 2 | - | - | - | - | - | - | - | 110 | 46 | 33 | 2 | 29 |
| | S1.3 QLD | 95 | 0 | 34 | 72 | 16 | 61 | 0.2 | 0.1 | 0.4 | 0.2 | 16 | 345 | - | - | - | - | - | - | - | 91 | 47 | 23 | 2 | 19 |
| | S1.4 SA | 23 | 0 | 9 | 20 | 4 | 13 | 0.0 | 0.0 | 0.0 | 0.0 | 4 | 123 | - | - | - | - | - | - | - | 101 | 48 | 31 | 2 | 21 |
| | S1.5 TAS | 14 | 0 | 0.0 | 0.0 | 1 | 5 | 2 | 10 | 0.0 | 0.0 | 0.4 | 2 | - | - | - | - | - | - | - | 51 | 46 | 2 | 1 | 2 |
| | S1.6 VIC | 98 | 0 | 23 | 43 | 27 | 105 | 2 | 1 | 0.1 | 0.0 | 15 | 707 | - | - | - | - | - | - | - | 114 | 49 | 36 | 3 | 26 |
| | S1.7 WA (Perth) | 47 | 0 | 19 | 39 | 11 | 45 | 0.0 | 0.0 | 0.0 | 0.0 | 10 | 342 | - | - | - | - | - | - | - | 124 | 47 | 39 | 2 | 37 |
| | NEM summary **a** | 346 | 0 | 108 | 223 | 63 | 239 | 7 | 13 | 1 | 0.3 | 55 | 1709 | - | - | - | - | - | - | - | 96 | 48 | 28 | 2 | 18 |
| | 7 Grids summary **b** | 393 | 0 | 127 | 262 | 74 | 284 | 7 | 13 | 1 | 0.3 | 65 | 2052 | - | - | - | - | - | - | - | 99 | 48 | 29 | 2 | 20 |
| Super Grid | S2.1 NEM | 346 | 6 | 80 | 168 | 65 | 249 | 7 | 13 | 1 | 0.3 | 46 | 637 | 0 | 21 | 25 | 24 | 0 | 0 | 2 | 83 | 48 | 14 | 9 | 13 |
| | S2.2 NEM+WA | 393 | 8 | 122 | 256 | 64 | 246 | 7 | 12 | 1 | 0.2 | 52 | 625 | 0 | 23 | 26 | 24 | 0 | 10 | 2 | 86 | 48 | 13 | 10 | 16 |
| | S2.3 NEM+NT | 346 | 6 | 96 | 203 | 57 | 217 | 7 | 12 | 1 | 0.3 | 47 | 611 | 0 | 19 | 29 | 25 | 1 | 0 | 2 | 84 | 48 | 14 | 9 | 13 |
| | S2.4 NEM+NT, WA | 393 | 8 | 99 | 209 | 78 | 302 | 7 | 12 | 1 | 0.2 | 49 | 467 | 0 | 18 | 40 | 23 | 2 | 10 | 2 | 86 | 47 | 11 | 11 | 17 |
| | S2.5 NEM+FNQ | 350 | 7 | 59 | 121 | 64 | 250 | 7 | 18 | 1 | 1 | 42 | 614 | 15 | 21 | 10 | 21 | 0 | 0 | 2 | 75 | 47 | 13 | 9 | 6 |
| | S2.6 NEM+FNQ, WA | 397 | 10 | 60 | 124 | 77 | 298 | 7 | 17 | 1 | 1 | 45 | 660 | 21 | 22 | 21 | 22 | 0 | 9 | 2 | 77 | 47 | 12 | 11 | 7 |
| | S2.7 NEM+FNQ, NT | 350 | 7 | 63 | 132 | 59 | 237 | 7 | 18 | 1 | 1 | 41 | 601 | 21 | 17 | 9 | 20 | 4 | 0 | 2 | 74 | 46 | 13 | 9 | 6 |
| | S2.8 NEM+FNQ, NT & WA | 397 | 12 | 60 | 122 | 73 | 294 | 7 | 18 | 1 | 1 | 39 | 678 | 22 | 20 | 24 | 18 | 10 | 9 | 2 | 76 | 47 | 12 | 12 | 6 |
| Smart Grid | S3.1 NEM | 346 | 4 | 93 | 197 | 60 | 229 | 7 | 12 | 1 | 0.3 | 38 | 304 | 0 | 19 | 15 | 33 | 0 | 0 | 2 | 78 | 47 | 9 | 8 | 13 |
| | S3.2 NEM+WA | 393 | 7 | 91 | 188 | 78 | 300 | 7 | 12 | 1 | 0.2 | 37 | 330 | 0 | 18 | 30 | 29 | 0 | 14 | 2 | 81 | 48 | 8 | 11 | 14 |
| | S3.3 NEM+NT | 346 | 5 | 79 | 165 | 65 | 247 | 7 | 12 | 1 | 0.2 | 37 | 403 | 0 | 19 | 14 | 32 | 2 | 0 | 2 | 78 | 48 | 10 | 8 | 12 |
| | S3.4 NEM+NT, WA | 393 | 8 | 50 | 105 | 93 | 356 | 7 | 13 | 1 | 0.3 | 49 | 481 | 0 | 22 | 31 | 27 | 4 | 11 | 2 | 81 | 48 | 11 | 11 | 11 |
| | S3.5 NEM+FNQ | 350 | 7 | 59 | 123 | 62 | 244 | 7 | 18 | 1 | 1 | 34 | 310 | 17 | 17 | 8 | 25 | 0 | 0 | 2 | 70 | 47 | 8 | 9 | 6 |
| | S3.6 NEM+FNQ, WA | 397 | 9 | 49 | 102 | 77 | 310 | 7 | 18 | 1 | 1 | 35 | 455 | 18 | 20 | 14 | 20 | 0 | 10 | 2 | 70 | 46 | 9 | 10 | 5 |
| | S3.7 NEM+FNQ, NT | 350 | 7 | 52 | 109 | 65 | 256 | 7 | 18 | 1 | 1 | 31 | 297 | 16 | 16 | 12 | 23 | 4 | 0 | 2 | 69 | 47 | 8 | 9 | 5 |
| | S3.8 NEM+FNQ, NT & WA | 397 | 10 | 48 | 101 | 77 | 310 | 7 | 18 | 1 | 1 | 31 | 414 | 21 | 22 | 17 | 19 | 7 | 10 | 2 | 70 | 46 | 8 | 11 | 5 |

Note. Energy demand is operational, which means rooftop PV is not included in the LCOE calculation, neither in the cost (numerator) nor in the energy (denominator) components.

[a] Here the National Electricity Market (NEM) is defined as a fully integrated electricity market including NSW, QLD, SA, TAS and VIC, but excluding FNQ, NT and WA.

[b] Mt Isa, which is located in FNQ and currently off-the-grid (energy consumption: 4.2 TWh p.a.), is connected to the main network only in the Super Grid and Smart Grid scenarios.

### 3.1 Energy affordability

As shown in Fig. 5 and Table A, in the 7 Grids scenario (S1.1-S1.7), the levelised cost of electricity (LCOE) ranges from $51-124/MWh in QLD, NSW, NT, SA, TAS, VIC and WA with a volume-weighted average of $99/MWh. The storage requirements are 65 GW, 2052 GWh in total. By contrast, in the Super Grid scenario (S2.1-S2.8), the LCOE of an integrated National Electricity Market (including QLD, NSW, SA, TAS and VIC) is in the range $74-86/MWh depending on whether connections to FNQ, NT and WA are built. In particular, a fully integrated energy system of NEM + FNQ, NT, WA in S2.8 costs only $76/MWh, which represents a reduction of $23/MWh in LCOE when compared to the volume-weighted average in the 7 Grids scenario. Of the HVDC extensions to FNQ, NT and WA, the NEM-FNQ link has the most significant influence on the reduction of LCOE in the NEM, as it provides access to a less-correlated wind resource in the Far North (17-19 degrees of latitude). In comparison, the NEM-NT and NEM-WA links primarily help to reduce the costs in NT (from $110/MWh in S1.2 to $84/MWh in S2.3) and WA (from $124/MWh in S1.7 to $86/MWh in S2.2). Accordingly, the total storage requirements in the Super Grid scenario decrease to 467-678 GWh - equivalent to the storage requirement in a single state, NSW or VIC, in the 7 Grids scenario. With flexible charging of electric cars, the LCOE in the Smart Grid scenario reduces further still to $69-81/MWh and the total storage requirements range from 297-481 GWh. A fully integrated energy system of NEM + FNQ, NT, WA in S3.8 costs $70/MWh and requires 31 GW, 414 GWh of storage capacity.

The LCOE comprises 4 components: (i) the levelised cost of generation (LCOG) which refers to the cost of energy sourced from solar, wind, existing hydro and bio; (ii) the storage; (iii) the transmission; and (iv) occasional energy spillage & loss. The levelised cost of balancing (LCOB) is made up of the storage, transmission and energy spillage & loss components, and is summed together with the LCOG to find the LCOE. As shown in Table 2, while the LCOG are similar across the scenarios ($46-49/MWh), large differences in the LCOB can be observed which reflect the characteristics of each scenario. For example, compared with the 7 Grids scenario, the Super Grid scenario features higher costs of transmission ($9-12/MWh) due to the construction of the HVDC backbone. However, the storage and energy spillage & loss components in the Super Grid scenario decrease substantially, which leads to large reductions in the LCOE as a whole.

The results also suggest that large-scale grid interconnection in the Super Grid scenario and the introduction of demand-side participation in the Smart Grid scenario help to reduce the cost of 100% renewables by 13-25/MWh and 18-30/MWh respectively. This is significant: each dollar of decrease in the LCOE equates to $400 million of cost savings in the energy industry per year. In fact, as shown in Table 2, in the Super Grid and Smart Grid scenarios, the installed capabilities of solar PV, wind and storage (both GW and GWh) are remarkably reduced when compared with the capacities in the 7 Grids scenario due to the benefits of wide geographical dispersion of solar and wind energy over a large area and the flexibility of electric car charging. Significantly, in the Super Grid and Smart Grid scenarios, the costs of 100% renewables including energy generation, storage and transmission can be competitive with current electricity prices in the Australian wholesale energy market ($80-110/MWh on average in 2019 [31]) and are lower than the cost of new-build coal and gas power stations (greater than $80/MWh) in Australia [26].

### 3.2 Energy security and reliability

A snapshot of the load profiles and generation mix is included in Fig. 6. As illustrated, in each of the 3 hypothetical 100% renewables scenarios, energy storage is responsible for large-scale energy shifting which is the key to energy supply and demand balance with high penetrations of solar and wind energy. In addition to large-scale energy time-shifting, STORES and DES also contribute to a variety of ancillary services such as frequency control and black start capability, which both help to build the resilience of the energy system and deliver a high level of energy security and reliability. The energy supply-demand balance data for the entire simulated period (10 years with 175,344 half-hourly intervals) are available from this Dropbox link: https://www.dropbox.com/sh/i21fhvy5tjgbsl9/AACAP72LKNE9h-GoVCSleTK6a?dl=0.

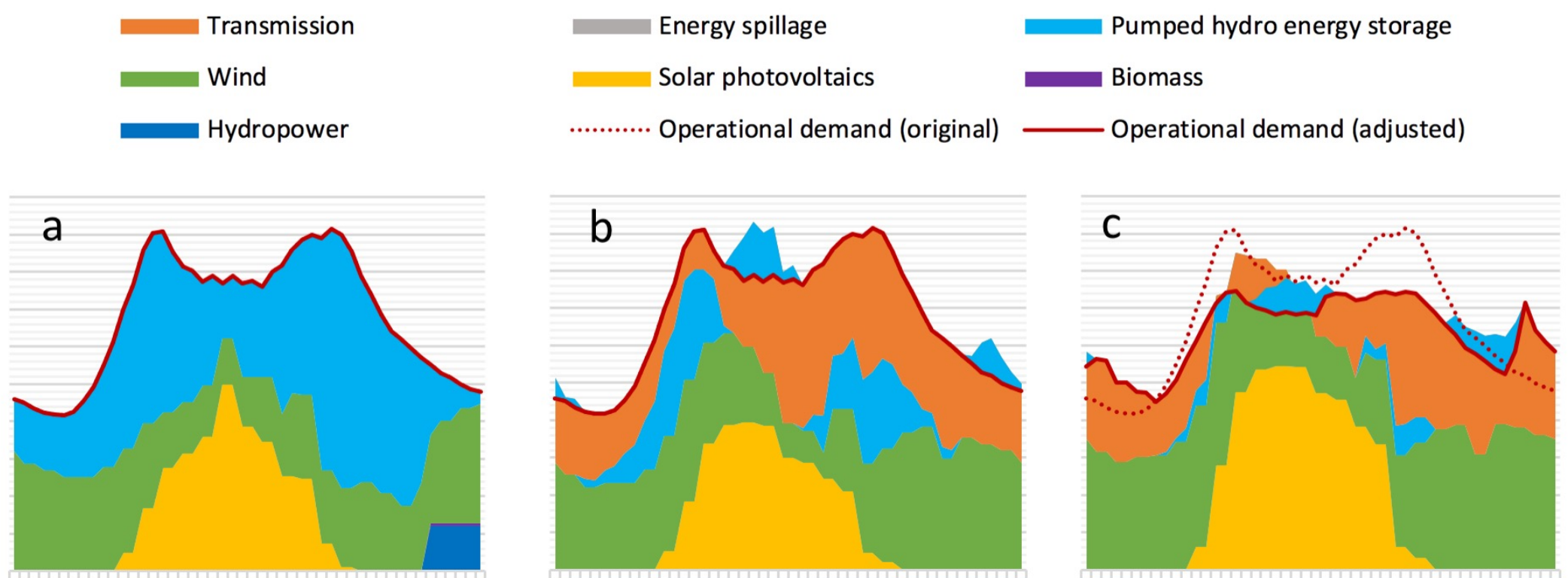

Fig. 6 Load profiles and generation mix for a day with low availability of wind energy in New South Wales in the 7 Grids (a), Super Grid (b) and Smart Grid (c) scenarios. Fig. 6-b and 6-c represent a fully integrated NEM + FNQ, NT and WA electricity system in the Super Grid (S2.8) and the Smart Grid (S3.8) scenarios respectively.

## 4. Discussion

In this study, we demonstrated that a zero-carbon, reliable and affordable electrical energy system that reduces Australian greenhouse gas emissions by 80% can be built based on: (i) solar photovoltaics, wind turbines, existing hydropower and biomass for power generation; (ii) pumped hydro (off-river) and electric car batteries for energy storage; and (iii) high-voltage direct-current and alternating-current for electricity transmission. At the end of 2019, the global installed capacities of these energy technologies were: solar photovoltaics 580 GW [1], wind turbines 623 GW [1], pumped hydro 181 GW [21], high-voltage direct current > 200 GW [32], and the deployment of electric cars had reached 7.2 million worldwide [33]. High levels of energy reliability and affordability were achieved through a synergy of flexible energy sources; interconnection of electricity grids over large areas; response from demand-side participation; and mass energy storage.

The novel features of this study include:

- Modelling of short-term off-river energy storage and distributed energy storage as two vast and low-cost energy storage technologies.
- Emphasis on large-area (million square kilometres) high-voltage transmission interconnection;
- Dependence on energy technologies already in very large-scale production;

- Analysis of Australia as an example Sunbelt country, where three quarters of the global population resides, and which lacks both seasonality of solar resources and cold winters;

Australia is located at low-moderate latitudes (the "Sunbelt") and has a tropical climate in many parts of the country. Long-term, seasonal energy storage requirements are low compared with Europe, North America and Northeast Asia. Hence, short-term off-river energy storage and distributed energy storage are ideal solutions for balancing variable solar and wind energy on timescales of minutes, hours and days. This strategy is highly likely to be transferable to other countries in the Sunbelt.

Deep decarbonisation of the energy sector, cost sensitivity and electric vehicles modelling are further discussed in Section 4.1-4.3.

**4.1 Deep decarbonisation of the energy sector**

In this work, we modelled a fully decarbonised electricity system in Australia along with electrification of heating, transportation and industry. Electrification of the energy sector is straightforward through the deployment of electric vehicles, electric heating, and electric equipment in place of fossil fuels as noted in Fig. 4. Most of these energy technologies are commercially available today and have already been deployed on a large scale worldwide. With rapid deployment in the near future, the cost and performance of these technologies would be further improved, and therefore become even more attractive as substitutes for existing fossil fuels-based energy technologies.

It is noted that electrification is not the only pathway to deep decarbonisation of the energy sector. For example, solar thermal energy can be collected through solar hot water systems (low temperature heat) or concentrating solar collectors for power generation (high temperature heat). However, in light of the rapidly declining cost of solar photovoltaics and the advantages of electric applications, such as high efficiency of energy conversion and current cost parity, we believe that renewable electricity from solar photovoltaics and wind turbines is likely to be dominant in high renewable energy futures. In other words, 100% renewable electricity and 100% renewable energy (including transport, heating and industry) may converge in the future energy systems.

Aviation and shipping are not included in the modelling. Direct electrification of aviation and shipping is difficult when compared with that of other transport modes, because they are more sensitive to the energy density (gravimetric and volumetric) of alternative transport fuels. While battery storage with electric motors could be a practical solution for short-haul flights and ships, the energy technologies to electrify long-haul flights and ships are still being developed. Nevertheless, renewables-based fuels such as hydrogen, ammonia and synthetic hydrocarbons via water electrolysis and chemical synthesis present a promising prospect for zero-carbon alternatives to jet fuel and heavy oil fuel.

In addition, Australia has relatively small emissions per capita from industry because Australian manufacturing of items such as iron and steel, cement and plastics is a relatively small fraction of the economy. Full electrification of industry in some other countries will be a relatively larger endeavour than in Australia. Production of synthetic hydrocarbon fuels (such as jet fuel) and plastics requires a sustainable source of carbon, which will probably necessitate carbon capture from the air.

**4.2 Cost sensitivity**

Modelling input and assumptions were further examined in a sensitivity analysis, where the values of cost components in the scenario S3.8 were varied between +25% and -25%. The levelised cost of electricity (LCOE) was most sensitive to changes in the cost of wind turbines and the discount rate. For example, a 25% increase in the discount rate led to a $9 increase in the LCOE. By contrast, the LCOE was less sensitive to the costs of solar photovoltaics, high-voltage direct-current transmission, energy storage, existing hydropower and biomass, and high-voltage alternating-current transmission. Given that the energy technologies included in this model have already been deployed on a large scale worldwide, the costs are well-known, and so we believe that these cost estimates for renewable energy systems are robust compared with technologies that are under research & development or in the demonstration stage.

**4.3 Electric vehicles modelling**

We note that demand-side participation in the Smart Grid scenario only refers to the flexibility of electric cars that charge in response to energy sufficiency in the system. Our modelling shows that the key requirement to effective use of car batteries to help meet demand is to avoid charging the batteries during morning and evening peak periods that last for a few hours each. Provided this criterion is met, the actual charging pattern matters little.

Vehicle-to-grid (V2G) technology was not included, though V2G could further lower the levelised cost of electricity through peak shaving. The impact of V2G on lithium-ion batteries is being investigated. On one hand, V2G has been linked with accelerated degradation, but others believe that the degradation of battery capacity and power output due to extensive charging and discharging operations can be effectively minimised through careful management of vehicle charging and discharging [34]. A future study to explore the benefits and challenges of utilising V2G technology will explore this opportunity. Additionally, in this model, millions of electric car batteries were aggregated and modelled as a "giant" battery. This provided a rough estimate of the benefit of integrating demand flexibility. A representation of distributed energy resources including hot water storage and household batteries with a high level of granularity would be included in future studies.

## Acknowledgements

This project has been supported by the Australian Government through the Australian Renewable Energy Agency (ARENA) and the Australian Centre for Advanced Photovoltaics (ACAP). Responsibility for the views, information or advice expressed herein is not accepted by the Australian Government. Thanks also to the Australian National University (ANU) Grand Challenge Programme *Zero-Carbon Energy for the Asia-Pacific* for funding. This work was supported by computational resources provided by the Australian Government through the National Computational Infrastructure facility under the ANU Merit Allocation Scheme.

## References

[1] International Renewable Energy Agency. Renewable capacity statistics 2020. 2020.
[2] Jacobson MZ, Delucchi MA, Bazouin G, Bauer ZAF, Heavey CC, Fisher E, et al. 100% clean and renewable wind, water, and sunlight (WWS) all-sector energy roadmaps for the 50 United States. Energ Environ Sci. 2015;8:2093-117.
[3] Jacobson MZ, Delucchi MA, Bauer ZAF, Goodman SC, Chapman WE, Cameron MA, et al. 100% Clean and Renewable Wind, Water, and Sunlight All-Sector Energy Roadmaps for 139 Countries of the World. Joule. 2017;1:108-21.
[4] Connolly D, Lund H, Mathiesen BV. Smart Energy Europe: The technical and economic impact of one potential 100% renewable energy scenario for the European Union. Renew Sust Energ Rev. 2016;60:1634-53.
[5] Lund H, Ostergaard PA, Connolly D, Mathiesen BV. Smart energy and smart energy systems. Energy. 2017;137:556-65.
[6] Ram M, Bogdanov D, Aghahosseini A, Gulagi A, Oyewo SA, Child M, et al. Global energy system based on 100% renewable energy – power, heat, transport and desalination sectors. LUT University; 2019.
[7] Bogdanov D, Farfan J, Sadovskaia K, Aghahosseini A, Child M, Gulagi A, et al. Radical transformation pathway towards sustainable electricity via evolutionary steps. Nat Commun. 2019;10.
[8] Blakers A, Stocks M, Lu B, Cheng C, Stocks R. Pathway to 100% Renewable Electricity. Ieee J Photovolt. 2019;9:1828-33.
[9] Baldwin K, Blakers A, Stocks M. Australia's renewable energy industry is delivering rapid and deep emissions cuts. Australian National University; 2018.
[10] Solargis. Global solar atlas 2.0 https://globalsolaratlas.info/download/world.
[11] Australian Department of the Environment and Energy. Quarterly update of Australia's national greenhouse gas inventory: June 2019. 2019.
[12] Australian Energy Market Operator. Integrated system plan for the National Electricity Market. 2018.
[13] Australian Bureau of Meteorology. Australian hourly solar irradiance gridded data http://www.bom.gov.au/climate/how/newproducts/IDCJAD0111.shtml.
[14] Australian bureau of Meteorology. Data services: weather station directory http://www.bom.gov.au/climate/data/stations/.
[15] Lu B, Blakers A, Stocks M. 90-100% renewable electricity for the South West Interconnected System of Western Australia. Energy. 2017;122:663-74.
[16] Geoscience Australia. Australian energy resource assessment: second edition. 2018.
[17] Australian Clean Energy Council. Clean energy australia report 2019. 2019.
[18] Gernaat DEHJ, Bogaart PW, van Vuuren DP, Biemans H, Niessink R. High-resolution assessment of global technical and economic hydropower potential. Nat Energy. 2017;2.
[19] Popp J, Lakner Z, Harangi-Rakos M, Fari M. The effect of bioenergy expansion: Food, energy, and environment. Renew Sust Energ Rev. 2014;32:559-78.
[20] World Nuclear Association. World Nuclear Power Reactors & Uranium Requirements https://www.world-nuclear.org/information-library/facts-and-figures/world-nuclear-power-reactors-and-uranium-requireme.aspx. 2020.

[21] U.S. Department of Energy. DOE Global Energy Storage Database https://www.energystorageexchange.org/projects/data_visualization.
[22] Australian Department of the Environment and Energy. National greenhouse gas inventory - Kyoto Protocol classifications https://ageis.climatechange.gov.au.
[23] Australian Energy Market Operator. Market data Western Australia: load summary http://data.wa.aemo.com.au/#load-summary.
[24] Australian Department of the Environment and Energy. Australian energy update 2019. 2019.
[25] Beyond Zero Emissions. Zero carbon industry plan: electrifying industry. 2018.
[26] Graham P, Hayward J, Foster J, Story O, Havas L. GenCost 2018: Updated projections of electricity generation technology costs. Commonwealth Scientific and Industrial Research Organisation; 2018.
[27] Blakers A, Stocks M, Currie J, Ho H, Hart P, Fulton R, et al. A public cost model for off-river pumped hydro. Australian National University; 2018.
[28] U.S. Energy Information Administration. Assessing HVDC transmission for impacts of non-dispatchable generation. 2018.
[29] Tamblyn J. Feasibility of a second Tasmanian interconnector. 2017.
[30] Australian Energy Market Operator. 100 per cent renewable study. 2013.
[31] Australian Energy Market Operator. National Electricity Market: data dashboard https://www.aemo.com.au/Electricity/National-Electricity-Market-NEM/Data-dashboard.
[32] International Energy Agency. IEA ETSAP technology brief E12: electricity transmission and distribution. 2014.
[33] International Energy Agency. Global EV outlook 2020 https://www.iea.org/reports/global-ev-outlook-2020. 2020.
[34] Uddin K, Jackson T, Widanage WD, Chouchelamane G, Jennings PA, Marco J. On the possibility of extending the lifetime of lithium-ion batteries through optimal V2G facilitated by an integrated vehicle and smart-grid system. Energy. 2017;133:710-22.

## Supplementary Information. Modelling assumptions for transport and heating electrification

Table SI-1, SI-2 include a summary of the modelling assumptions for transport and heating electrification. Energy use (petajoule, PJ) and electricity demand (terawatt-hour, TWh) show the total consumption of fossil fuels or electricity in New South Wales, the Northern Territory (Alice Springs), Queensland, South Australia, Tasmania, Victoria and Western Australia (Perth) in 2017-18. Alice Springs and Perth were assumed to constitute 10% and 95% of transport and heating demand in the Northern Territory and Western Australia, respectively.

Table SI-1. Assumptions for the electrification of land transport (excluding aviation and shipping)

| | | Passenger car | Light commercial vehicle | Rigid truck | Articulated truck | Non-freight truck | Bus | Motorcycle | Rail |
|---|---|---|---|---|---|---|---|---|---|
| **1** | **Transport** | | | | | | | | |
| 1.1 | Number of motor vehicles (million) | 14.3 | 3.2 | 0.5 | 0.1 | 0.02 | 0.08 | 0.86 | - |
| 1.2 | Average travel distance (1,000 km) | 12.6 | 16.4 | 20.8 | 79.4 | 13.1 | 26.9 | 2.6 | - |
| 1.3 | Average fuel consumption (L/100 km) | 10.8 | 12.5 | 28.6 | 55.2 | 21.3 | 28.4 | 5.8 | - |
| 1.4 | Hourly/half-hourly travel pattern | Sydney GMR HTS | Sydney GMR HTS | NSW RMS | NSW RMS | NSW RMS | PTV | Sydney GMR HTS | Flat |
| **2** | **Electric vehicles** | | | | | | | | |
| 2.1 | Energy consumption (kWh/100 km) | 27 | 32 | 80 | 160 | 73 | 76 | 14 | - |
| 2.2 | Vampire loss (per day) | 1% | 1% | 1% | 1% | 1% | 1% | 1% | - |
| 2.3 | Transmission & distribution loss | 7.5% | 7.5% | 7.5% | 7.5% | 7.5% | 7.5% | 7.5% | 7.5% |
| 2.4 | Charging | Levels 1 & 2 | Levels 1 & 2 | Levels 1 & 2 | Levels 1 & 2 | Levels 1 & 2 | Levels 1 & 2 | Levels 1 & 2 | Electric traction |
| 2.5 | Charging efficiency | 85% | 85% | 85% | 85% | 85% | 85% | 85% | - |
| **3** | **Grid integration** | | | | | | | | |
| 3.1 | Electricity demand (TWh p.a.) | 50.6 | 17.2 | 8.7 | 12.8 | 0.22 | 1.9 | 0.33 | 5.1 |
| 3.2 | Demand flexibility | 80% | 0% | 0% | 0% | 0% | 0% | 0% | 0% |
| 3.3 | Minimum reserve | 25% | - | - | - | - | - | - | - |

Abbreviations: Sydney Greater Metropolitan Region Household Travel Survey (Sydney GMR HTS), New South Wales Roads and Maritime Services (NSW RMS), Public Transport Victoria (PTV).

Table SI-2. Assumptions for the electrification of heating in residential and commercial buildings

| | | Space heating | Water heating | Cooking |
|---|---|---|---|---|
| **1** | **Heating: natural gas and liquefied petroleum gas** | | | |
| 1.1 | Energy use (PJ) | 23 | 22 | 8 |
| 1.2 | Average fuel efficiency | 75-90% | 67% | 40% |
| 1.3 | Hourly distribution | Temperature & occupancy | Temperature & occupancy | Temperature & occupancy |
| **2** | **Heat pump and electric appliance** | | | |
| 2.1 | Average power rating (kW) | 4 | 1.5 | 1.5-2.5 |
| 2.2 | Coefficient of performance/efficiency | 200%-600% | 200%-600% | 70% |
| 2.3 | Average tank size (L) | - | 250-350 | - |
| 2.4 | Heat loss (per day) | - | 2 kWh | - |
| 2.5 | Transmission & distribution loss | 7.5% | 7.5% | 7.5% |
| **3** | **Grid integration** | | | |
| 3.1 | Electricity demand (TWh p.a.) | 8.1 | 6.6 | 4.9 |
| 3.2 | Demand flexibility | - | 0% | - |